\documentclass[%
 superscriptaddress,
 amsmath,amssymb,
 aps,
prb,
twocolumn
]{revtex4-1}

\usepackage{graphicx}
\usepackage{dcolumn}
\usepackage{bm}
\usepackage{siunitx} 

\begin{document}

\preprint{APS/123-QED}

\title{Nonharmonic Phonons in $\alpha$-Iron at High Temperatures}

\author{L. Mauger} 
\affiliation{
California Institute of Technology, W. M. Keck Laboratory 138-78,
Pasadena, CA 91125, USA
}

\author{M. S.~Lucas}
\affiliation{Air Force Research Laboratory, Wright-Patterson AFB, OH 45433, USA}

\author{J. A. Mu{\~n}oz}
\affiliation{
California Institute of Technology, W. M. Keck Laboratory 138-78,
Pasadena, CA 91125, USA
}

\author{S. J. Tracy}
\affiliation{
California Institute of Technology, W. M. Keck Laboratory 138-78,
Pasadena, CA 91125, USA
}

\author{M. Kresch}
\affiliation{Marft, Inc. 101 Irene Court, Belmont, CA 94002, USA}

\author{Yuming Xiao}
\affiliation{
HPCAT, Geophysical Laboratory, Carnegie Institute of Washington, Argonne, IL 60439, USA
}

\author{Paul Chow}
\affiliation{
HPCAT, Geophysical Laboratory, Carnegie Institute of Washington, Argonne, IL 60439, USA
}

\author{B. Fultz}
\affiliation{
California Institute of Technology, W. M. Keck Laboratory 138-78,
Pasadena, CA 91125, USA
}

\date{\today}

\begin{abstract}

Phonon densities of states (DOS) of bcc $\alpha$-$^{57}$Fe were measured from room temperature through the 1044K Curie transition and the 1185K fcc $\gamma$-Fe phase transition using nuclear resonant inelastic x-ray scattering. 
At higher temperatures all phonons shift to lower energies (soften) with thermal expansion, but the low transverse modes soften especially rapidly above 700K, showing strongly nonharmonic behavior that persists through the magnetic transition.  
Interatomic force constants for the bcc phase were obtained by iteratively fitting a Born-von K\'{a}rm\'{a}n model to the experimental phonon spectra using a genetic algorithm optimization.  
The second-nearest-neighbor fitted axial force constants weakened significantly at elevated temperatures.
An unusually large nonharmonic behavior is reported, which increases the vibrational entropy and accounts for a contribution of 35 meV/atom in the free energy at high temperatures.  
The nonharmonic contribution to the vibrational entropy follows the thermal trend of the magnetic entropy, and may be coupled to magnetic excitations. 
A small change in vibrational entropy across the $\alpha$-$\gamma$ structural phase transformation is also reported.   
\end{abstract}

\pacs{Valid PACS appear here}
\maketitle

\section{\label{sec:level1}Introduction}
In its metallic form, iron exhibits fascinating physics, plays a central role in geophysics, and is of paramount importance to metallurgy. 
Iron is polymorphic under temperature, pressure, and alloying, and both its magnetic properties and its mechanical properties undergo major changes with crystal structure. 
The thermodynamics of the temperature-induced polymorphism of iron have been of interest for many years. 
A proper thermodynamic treatment of metallic iron must consider the energetics as well as the degrees of freedom of electrons, phonons, and spins, and the couplings between them. 
Although this is a complex problem, it has received longstanding interest both for its own sake, and for predicting the phases of iron alloys with an eye to controlling them\cite{pepperhoff_constitution_2010,kormann_lambda_2014}.

There have been a large number of heat capacity\cite{white_thermophysical_1997,purdue_university_thermophysical_1970} and elastic constant\cite{dever_temperature_1972,rayne_elastic_1961} measurements of iron at various temperatures, and the thermodynamic entropy of iron is sufficiently reliable to be used in Calphad-type calculations of free energy \cite{jacobs_thermodynamic_2010,dinsdale_sgte_1991}. 
There have been a number of efforts to create predictive thermodynamics models by resolving the entropy into contributions from phonons, spins, and electrons \cite{weiss_components_1956,kormann_free_2008,jacobs_thermodynamic_2010}. 
Phonons make the largest contribution to the entropy at elevated temperatures, therefore the accuracy of the phonon entropy is critical. 
A harmonic model can account for most of the vibrational entropy of elemental solids.
The vibrational entropy of iron is quite large, however, exceeding 6 $k_{\rm B}$/atom at 1000K, so even errors of a few percent are thermodynamically important. 
The quasiharmonic model of vibrational entropy incorporates the phonon frequency shifts that result from finite temperature thermal expansion, but it neglects many other nonharmonic physical interactions.
Phonons interact through the cubic and quartic parts of the interatomic potential\cite{wallace_thermodynamics_1998}. 
These anharmonic phonon-phonon effects further change the phonon frequencies and shorten their lifetimes resulting in thermal broadening of phonon spectra\cite{kresch_neutron_2007,kresch_phonons_2008,kresch_temperature_2009}. 
Thermal excitations of electrons and magnons also affect the phonon frequencies through adiabatic electron-phonon and magnon-phonon interactions.
The impact of these physical interactions on the vibrational entropy and free energy has been shown to be important in many materials\cite{delaire_electron-phonon_2008,delaire_adiabatic_2008}, and their role in the vibrational thermodynamics of iron warrants further investigation.

Inelastic neutron scattering studies of phonon dispersions in iron provide essential information on the phonon contribution to entropy, and how it changes with temperature\cite{minkiewicz_phonon_1967,van_dijk__1968,satija_neutron_1985,zarestky_lattice_1987,neuhaus_phonon_1997}.
High temperature phonon dispersions show significantly decreased phonon frequencies with thermodynamic implications \cite{neuhaus_phonon_1997}.
These measurements also provide insight into the mechanism of the polymorphic transitions, and correlate with the inherent weaknesses of the bcc structure\cite{hasegawa_phonon_1987,petry_phonons_1989,heiming_are_1991,petry_dynamical_1995}.
However, the existing experimental results are somewhat sparse in temperature.

Ab initio investigations have attempted to identify individual contributions to the free energy of Fe and its alloys, but earlier studies relying on quasiharmonic approximations at high temperatures had limited success.  
Electronic structure calculations on iron have advanced considerably in the past few years, and recent work has carefully considered the different contributions of magnetism and vibrations to the thermodynamics of the bcc phase. 
Only recently have computational developments permitted DFT calculations to reproduce the observed high temperature phonon behavior by including the finite temperature magnetic configurations and electron-phonon coupling \cite{kormann_atomic_2012,kormann_rescaled_2010,leonov_calculated_2012,verstraete_ab_2013}.  
These developments suggest we may soon sort out the complex interactions in polymorphic iron, but experimental validation is still needed. 
Measurements of phonon dynamics through the Curie point at 1043K up to the $\gamma$-Fe phase transformation can provide further insight into the physical interactions and thermodynamics governing the complex behavior of iron. 

Here we report results of an experimental study of the vibrational properties of bcc $\alpha$-Fe at elevated temperatures, and an analysis of its interatomic interactions and thermodynamic functions.
Nuclear resonant inelastic x-ray scattering (NRIXS) was used to measure vibrational spectra of bcc Fe and obtain reliable phonon densities of states (DOS) which, unlike most phonon dispersion measurements, can be used directly in thermodynamic functions. 
A methodology was developed for reliably extracting the temperature-dependent interatomic force constants, and consequently phonon dispersions, from the phonon DOS spectra.
Much of the high temperature nonharmonic phonon dynamics depends on the rapid softening of the 2NN interatomic forces, and the resulting softening of transverse phonons in $\Gamma$-N direction. 
The vibrational entropy is assessed with different models for predicting high temperature thermodynamics.
We report a large nonharmonic contribution to the phonon entropy, and suggest that it originates with effects of magnetic excitations on the phonon spectra.

\section{\label{sec:level1}Experimental}

Nuclear resonant inelastic x-ray scattering (NRIXS) measurements were performed on bcc $\alpha$-Fe at high temperatures.  
NRIXS is a low background technique that provides direct access to the full phonon density of states (DOS)\cite{chumakov_experimental_1999,sturhahn_theoretical_1999}. 
NRIXS spectra were collected from a \SI{25}{\micro\meter} thick Fe foil of 99.9\% purity and 95\% $^{57}$Fe isotopic enrichment.
Measurements were performed at beamline 16ID-D of the Advanced Photon Source at Argonne National Laboratory using a radiative heating furnace. 
This NRIXS vacuum furnace used a narrow kapton window to permit the x-rays to access the sample. 
The Fe foil was either held by two Ta heat shields adjacent to a thermocouple, or mounted directly on the thermocouple.
The NRIXS measurements performed below room temperature employed a He flow Be-dome cryostat.
The temperatures were accurate to within $\pm$20K, where ambiguity comes from comparing the furnace thermocouple measurements to in-situ nuclear forward scattering and the NRIXS-derived detailed balance temperature calculations following the procedures described in literature\cite{sturhahn_conuss_2000,sturhahn_geophysical_2007}.

An avalanche photodiode was positioned at approximately 90$^{\circ}$ from the incident beam to collect re-radiated photons beginning approximately \SI{20}{\nano\second} after the synchrotron pulse. 
The incident photon energy was tuned to \SI{14.413}{\kilo\electronvolt} using a high-resolution silicon crystal monochromator to provide a narrow distribution of energies with a FWHM of \SI{2.3}{\milli\electronvolt}. The incident photon energy was scanned through a range of $\pm$\SI{120}{\milli\electronvolt}, centered on the nuclear resonant energy.
The Phoenix reduction package was used to extract phonon DOS spectra from the collected spectra\cite{sturhahn_conuss_2000}.
Lamb-M\"{o}ssbauer factors from this reduction are compared with literature values in Fig. \ref{fig:LambMossabuer} of the appendix.

\section{\label{sec:level1}Force Constant Analysis}
Many thermal properties of crystalline solids can be explained by a simple model of the crystal as a set of massive nuclei whose interactions act like springs, providing a restoring force against displacements. 
This model was developed by Born and von K\'{a}rm\'{a}n (BvK), and transforms the real space interatomic forces into a dynamical matrix\cite{wallace_thermodynamics_1998}. 
While this model is commonly employed for fitting phonon dispersions, its utility for fitting phonon DOS spectra is less straightforward. 
Phonon DOS spectra are an aggregate of all phonon modes in reciprocal space; therefore, fitting force constants to a phonon DOS spectrum is more challenging than modeling phonon dispersions. 
To model our phonon DOS spectra, trial force constants were used to construct a dynamical matrix, $\boldsymbol{D}(\boldsymbol{q})$, which was diagonalized at a randomly-distributed set of 3.375 million $q$-points in the first Brillouin zone to collect the spectrum of phonon frequencies, $\omega^{2}$,
\begin{equation}
\label{eq:DynamicalMatrix}
M\omega^{2}\boldsymbol{\epsilon} = \boldsymbol{D}(\boldsymbol{q})\boldsymbol{\epsilon}
\end{equation}
where $M$ is the mass of the atom and $\boldsymbol{\epsilon}$ is the polarization of the phonon mode corresponding to reciprocal space vector $\boldsymbol{q}$. 
This BvK model was embedded in a genetic algorithm global optimization framework, where trial sets of force constants were generated randomly according to the differential evolution algorithm and the resulting DOS are compared with experimental data \cite{storn_differential_1997,mckerns_building_2012,mckerns_mystic:_2009}.
Each NRIXS DOS was fit independently to obtain a force constant tensor that minimized the sum of squared differences between the model and the experimental phonon DOS.
The optimizations used populations of 50 members which ``evolve" until they converge, typically after a few hundred generations, on a set of force constants that gave the "best fit to the experimental NRIXS DOS.
These optimizations were repeated several times to ensure convergence. 
For the optimization process, the highest energy feature of each phonon DOS spectrum was fit to a Gaussian distribution, the distribution was then used to replace the high energy tails of each DOS used for fitting.
This was done to standardize the phonon cutoff energy across the data set and suppress fitting to higher energy noise (which results from the data reduction and would not contribute meaningfully to the optimizations).

The BvK optimization was tried in several different configurations, each permitting a different number of nearest-neighbor (NN) force constants to vary.
The largest optimizations included atomic interactions through 5NN (13 independent force constants) which is consistent with the number of variables commonly used to fit neutron triple axis dispersion data in this system \cite{minkiewicz_phonon_1967,van_dijk__1968,klotz_phonon_2000}. 
The fitting process was also performed with fewer variable nearest-neighbors forces (leaving more distant force constants fixed to 300K tensorial force constants from literature\cite{minkiewicz_phonon_1967}). 
The most restrictive case limited the dynamics to interactions through 2NN (4 tensorial force constants). 

To test our methodology, we calculated the phonon DOS using force constants from Minkiewicz, et al.,\cite{minkiewicz_phonon_1967} and convolved it with our NRIXS experimental resolution function.
This DOS was optimized using the genetic algorithm DOS fitting method, with the number of variable nearest-neighbor force constants ranging from 2NN to 5NN. 
The optimizations that included interactions through the 4NN shell reproduced the known force constants accurately across several test cases, although only slightly better than optimizations that included interactions through only the 2NN shells. 
It was found that allowing variations through the 5NN shell noticeably increased the error in the tensorial force constants found by the algorithm. 
Accordingly, the results presented here are from the optimizations where only the first 2NN force constants were allowed to vary, except in Fig. \ref{fig:ForceConstants} where values for fits through 4NN are included for comparison.

\section{\label{sec:level1}Results}
 
\subsection{\label{sec:level2}Phonons}

\begin{figure}[b]
\includegraphics[width=2.5in]{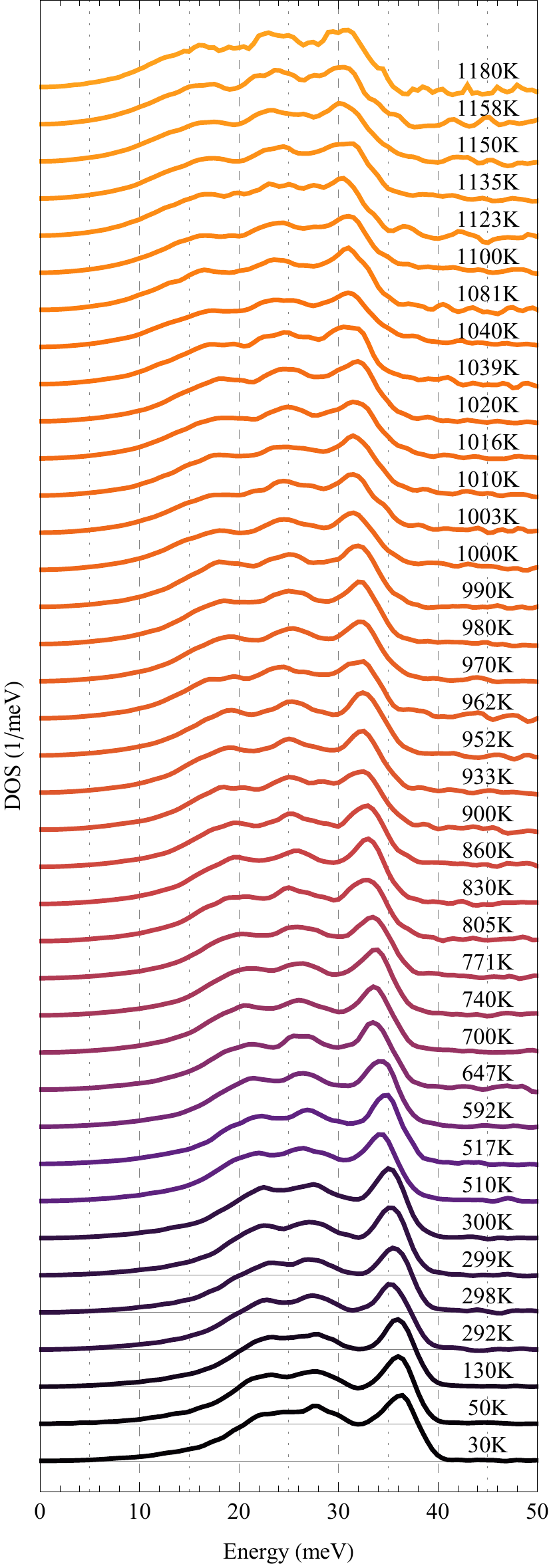}
\caption{\label{fig:DOS} The $^{57}$Fe phonon DOS extracted from NRIXS measurements at various temperatures. The spectra are normalized and offset for comparison.}
\end{figure}

\begin{figure}[b]
\includegraphics[width=3.25in]{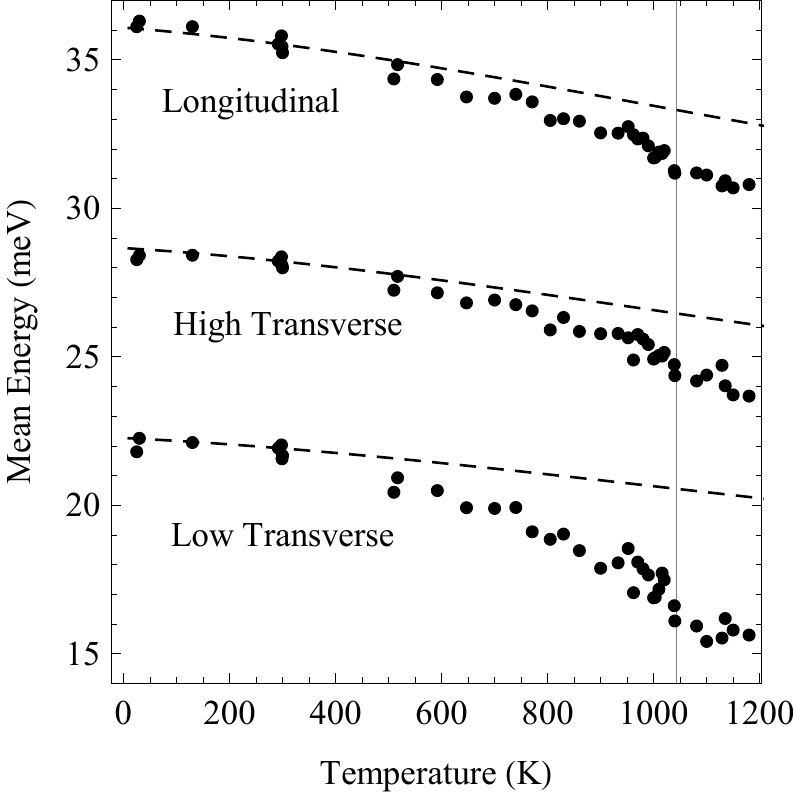}
\caption{\label{fig:ModeTracking} The measured NRIXS DOS were fit with three Lorentzian curves to find a characteristic mode energy for each phonon branch. The softening of these mode features is compared with a quasiharmonic model prediction from low temperature measurements (dashed lines). The Curie temperature at 1044K is marked by a vertical line.}
\end{figure}

The $^{57}$Fe phonon DOS spectra for $\alpha$-Fe from 30K to just below the $\gamma$-Fe transition at 1185K are shown in Fig.~\ref{fig:DOS}. 
All phonon modes shift to lower energies (soften) with increasing temperature, although some soften noticeably more than others. 
The phonon DOS of bcc Fe has three features corresponding to the Van Hove singularities of the longitudinal and two transverse acoustic phonon branches. 
The mean energies of the three features were obtained by simultaneously fitting three Lorentzian curves to the measured DOS spectra. 
The temperature dependence of these DOS features, displayed in Fig. \ref{fig:ModeTracking}, show that the low transverse phonons soften more than other modes.
While phonon softening with increasing temperature is ubiquitous in most materials, the large preferential softening of certain modes suggests strongly nonharmonic behavior of the lattice vibrations in $\alpha$-Fe at temperatures above 700K. 

\subsection{\label{sec:level2}Quasiharmonic Model}

The quasiharmonic model for predicting phonon frequencies employs the measured thermal expansion and a Gr\"{u}neisen parameter to account for how phonon frequencies deviate from the harmonic model at elevated temperatures. 
The quasiharmonic phonon frequencies, $\omega_{i}^{\rm qh}(T)$, are 
\begin{equation}
\label{omegaQH}
\omega_{i}^{\rm qh}(T) = \omega_{i}^{\rm 300K}(1-\overline{\gamma}_{\rm th}\frac{V_{T}-V_{\rm 300K}}{V_{\rm 300K}}).
\end{equation}
where $\omega_{i}^{\rm 300K}$ is the measured value of the $i^{th}$ phonon frequency at 300K,  $\overline{\gamma}_{\rm th}$ is the thermal Gr\"{u}neisen parameter and $V_{T}$ is the observed volume of the system at temperature $T$. 
This expression comes from the definition of the microscopic mode Gr\"{u}neisen parameter, $\gamma_{i}=(-\frac{\partial \ln\omega_{i}}{\partial \ln V})_{\rm T}\simeq-\frac{V}{\omega_{i}}\frac{\Delta\omega_{i}}{\Delta V}$,
where a thermal Gr\"{u}neisen parameter\cite{anderson_grueneisen_2000}, $\overline{\gamma}_{\rm th}$, is commonly used in the absence of detailed experimental observations of the mode Gr\"{u}neisen parameters, $\gamma_{i}$\cite{fultz_vibrational_2010}. 
The thermal Gr\"{u}neisen parameter can be calculated from observed bulk material properties; in the following analysis Anderson's value of 1.81 for $\alpha$-Fe is used\cite{anderson_grueneisen_2000}.
The quasiharmonic prediction from ambient temperature is shown by dashed lines in Fig. \ref{fig:ModeTracking} for each acoustic mode feature in the phonon DOS spectra.   
At temperatures beyond 800K the mean phonon energies for each acoustic branch soften more rapidly than predicted by the quasiharmonic model.
At 1180K, just before the $\gamma$-Fe structural transition, the average phonon energy has decreased by 19\% of its low temperature value, more than twice the 8\% softening predicted by the quasiharmonic model. 

A temperature-dependent thermal Gr\"{u}neisen parameter was also calculated from the observed bulk properties of $\alpha$-Fe using the expression
\begin{equation}
\label{eq:GammaMean}
\overline{\gamma}_{\rm th}(T)=\frac{\alpha (T) B_{\rm T}(T)\nu(T)}{C_{\rm V}(T)}
\end{equation}
where $B_{\rm T}(T)$ is the bulk modulus\cite{dever_temperature_1972,rayne_elastic_1961,adams_elastic_2006}, $\alpha(T)$ is the linear thermal expansion\cite{purdue_university_thermophysical_1970,liu_calibration_2004}, $\nu(T)$ is the crystalline volume per atom\cite{purdue_university_thermophysical_1970}, and $C_{\rm V}(T)$ is calculated by integrating the low temperature phonon spectra\cite{minkiewicz_phonon_1967, van_dijk__1968} or a Debye model \cite{weiss_components_1956}. 
The temperature-dependent Gr\"{u}neisen parameters that can be created by various combinations of physical constants from the literature range from 1.7 to 2.2 over the temperatures of interest.
However, including temperature-dependent parameters in our analysis did not significantly alter the quantitative results provided by the quasiharmonic model. 
Our further analysis therefore used the simpler approach with a constant thermal Gr\"{u}neisen parameter of 1.81 \cite{anderson_grueneisen_2000}. 

\subsection{\label{sec:level2}Vibrational Entropy}

\begin{figure}[b]
\includegraphics[width=3.3in]{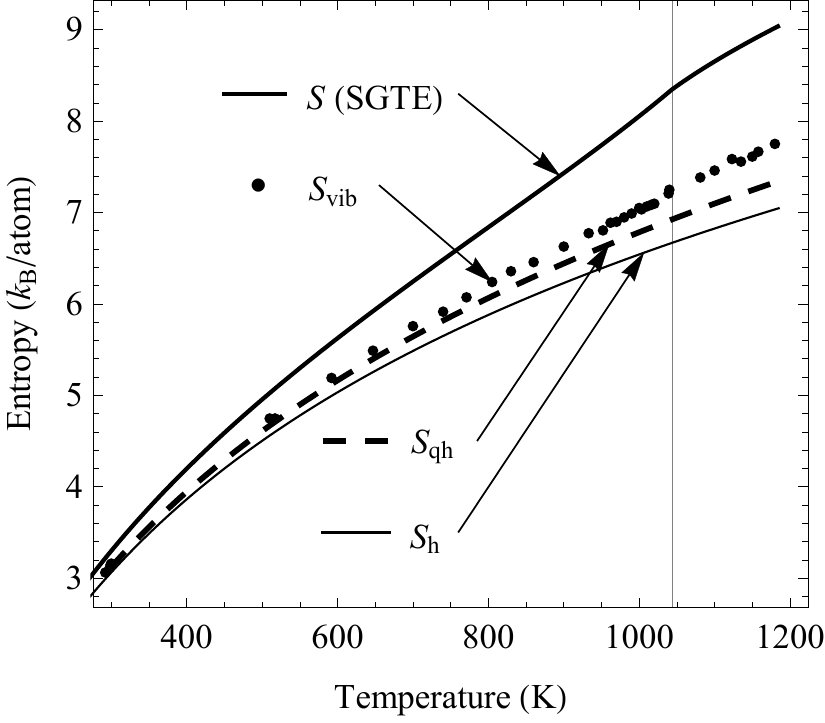}
\caption{\label{fig:VibrationalEntropy} Our measured vibrational entropy (points) compared with the Scientific Group Thermodata Europe total entropy (thick line), the quasiharmonic vibrational entropy estimate (dashed line), and the harmonic vibrational entropy (thin line).}
\end{figure} 

The total entropy of iron is often split into component entropies attributed to different physical phenomena,
\begin{equation}
\label{eq:totalEntropy}
S(T) = S_{\rm vib}(T) + S_{\rm el}(T) + S_{\rm mag}(T)
\end{equation}
where $S_{\rm vib}(T)$ is the vibrational entropy, $S_{\rm el}(T)$ is the electronic entropy, and $S_{\rm mag}(T)$ is the magnetic entropy. 
While this division neglects the complex interplay of excitations in real materials at elevated temperatures, it can still be useful for deconstructing thermodynamic models, especially when experimental observations focus on a subset of the physical interactions.
Accurate values of the vibrational entropy, $S_{\rm vib}(T)$, can be obtained directly from the experimentally measured phonon DOS as
\begin{equation}
\label{eq:VibrationalEntropy}
S_{\rm vib}(T)= 3k_{\rm B}\int g_{T}(E)\lbrace(n+1)\ln(n+1)-n\ln(n)\rbrace dE
\end{equation}
where $k_{\rm B}$ is the Boltzmann constant, $g_{T}(E)$ is the measured DOS at temperature, $T$, and $n$ is a Planck distribution evaluated at $T$, for a given energy, $E$. 
When experimental DOS spectra are available for a given temperature, this expression provides accurate entropy values that include both quasiharmonic effects and also nonharmonic effects (to first order)\cite{wallace_thermodynamics_1998}. 
The total vibrational entropies from NRIXS DOS spectra, $S_{\rm vib}$, are compared with the total entropy from the SGTE database, $S$, in Fig. \ref{fig:VibrationalEntropy}, together with the entropies of the harmonic and quasiharmonic models.
The total vibrational entropy, $S_{\rm vib}(T)$, can be divided into component entropies as
\begin{equation}
\label{VibEntropyHarmonicComponents}
S_{\rm vib}(T)=S_{\rm h}(T)+\Delta S_{\rm qh}(T)+\Delta S_{\rm nh}(T)
\end{equation}
where $S_{\rm h}(T)$ is harmonic vibrational entropy, $\Delta S_{\rm qh}(T) \equiv S_{\rm qh}(T)-S_{\rm h}(T)$ is the purely quasiharmonic contribution, and $\Delta S_{\rm nh}(T) \equiv S_{\rm vib}(T)-S_{\rm qh}(T)$, is the nonharmonic contribution. 
Figure \ref{fig:VibrationalEntropy} shows that both the harmonic model and the quasiharmonic model significantly underestimate the vibrational entropy obtained from NRIXS measurements. 
Above 1000K, the nonharmonic vibrational entropy, $\Delta S_{\rm nh}$, is larger than the quasiharmonic contribution, $\Delta S_{\rm qh}$.
At the highest temperatures the nonharmonic vibrational entropy, $\Delta S_{\rm nh}$, results in a 0.35$k_{\rm B}$/atom (5\%) increase over the vibrational entropy provided by the quasiharmonic model.

\subsection{\label{sec:level2}Born-von K\'{a}rm\'{a}n Fits}

\begin{figure}
\includegraphics[width=3.3in]{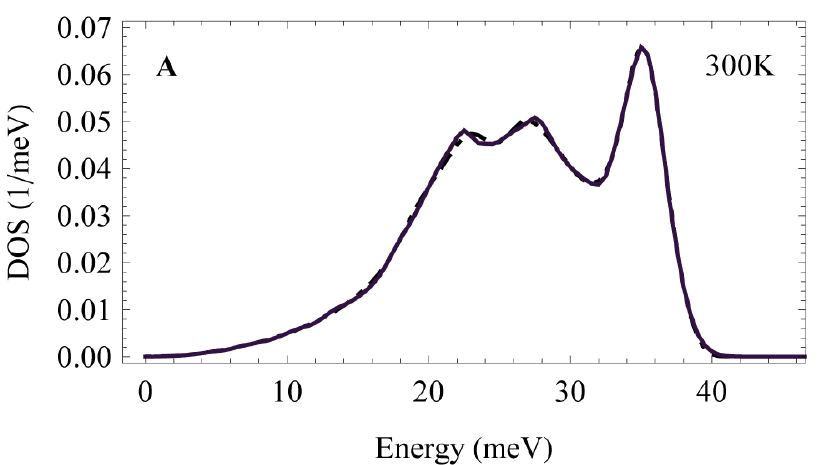}
\includegraphics[width=3.3in]{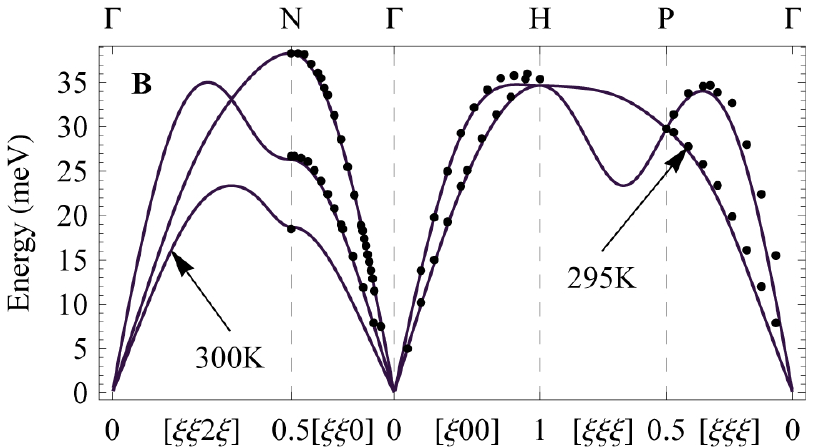}
\includegraphics[width=3.3in]{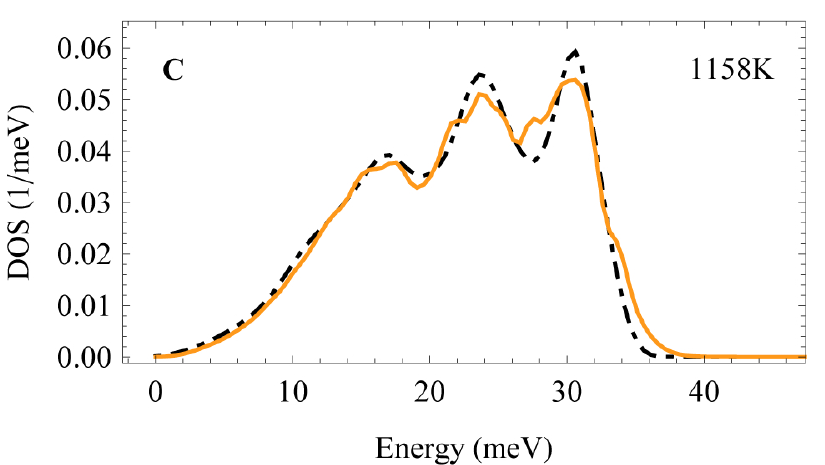}
\includegraphics[width=3.3in]{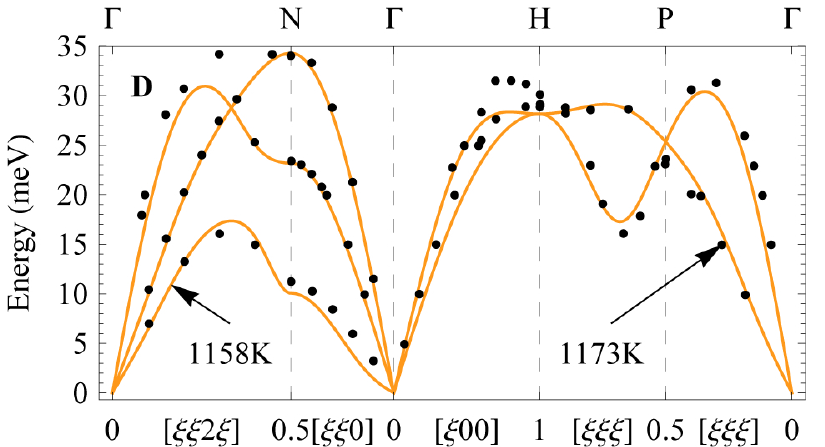}
\caption{\label{fig:ExampleFits} NRIXS DOS fits and corresponding BvK dispersions are compared with literature. Panels A and C compare the BvK fits (dashed lines) to the phonon DOS spectra at 300K and 1158K, respectively. Panel B compares our calculated 300K dispersions with the 295K neutron triple axis measurements of Minkiewicz, et al., (dots)\cite{minkiewicz_phonon_1967}. Panel D compares our calculated 1158K dispersions with the 1173K neutron triple axis measurements of Neuhaus, et al. (dots)\cite{neuhaus_phonon_1997}.}
\end{figure}

Tensorial force constants were optimized to fit each NRIXS DOS spectrum using the genetic evolution fitting algorithm, permitting the calculation of phonon dispersions at each temperature. 
Typical fits and dispersions are shown in Fig. \ref{fig:ExampleFits}.
The fitting procedure reproduces the DOS spectra quite well, and also generates phonon dispersions consistent with previous triple-axis neutron measurements\cite{neuhaus_phonon_1997,minkiewicz_phonon_1967}.   
\begin{figure}[b]
\includegraphics[width=3.3in]{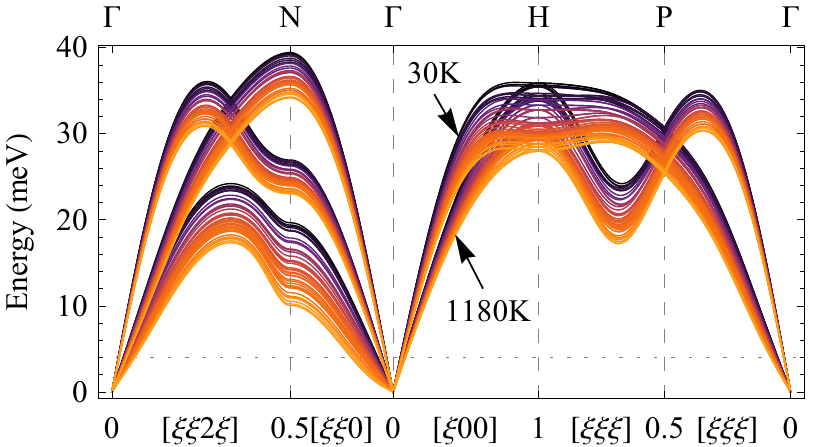}
\caption{\label{fig:Dispersions} 
Phonon dispersions resulting from force constant optimizations which permitted only 2NN force constants to vary (4 variables). Colors correspond to temperatures as labeled in Fig \ref{fig:DOS}.
}
\end{figure}
The calculated phonon dispersions corresponding to each NRIXS DOS measurement are displayed along the high symmetry directions and the $[\xi\xi2\xi]$ direction in Fig. \ref{fig:Dispersions}.
The phonon dispersions change monotonically with temperature, exhibiting significant softening at the highest temperatures, consistent with the phonon DOS spectra. 
These calculated dispersions show the trends identified by prior neutron scattering measurements\cite{neuhaus_phonon_1997}, and
elastic moduli extracted by fitting the low-$q$ portions of the dispersion branches are in good agreement with measured elastic moduli\cite{rayne_elastic_1961,dever_temperature_1972}. 
At elevated temperatures the optimized BvK fits began to segregate into two distinct solution basins.
The basins had similar energy spectra, but quite different phonon dispersions and force constants. 
The second basin of fits was characterized by H-point phonon energies that were significantly higher than the N-point longitudinal phonons.
They usually had lower qualities of fit than the primary basin, especially when the optimizations included higher nearest-neighbor interactions.
Fits from this second basin were easily identified as erroneous from their discontinuous changes in tensorial force constants with temperature, and their departure from measured dispersions\cite{neuhaus_phonon_1997}.
Thus, BvK fits from the second solution basin were excluded from further analysis.


The calculated dispersions demonstrate that some phonons soften significantly more than others. 
\begin{figure}[b]
\includegraphics[width=3.25in]{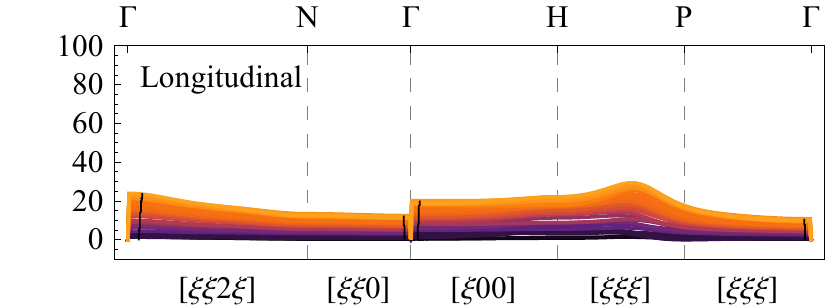}
\includegraphics[width=3.25in]{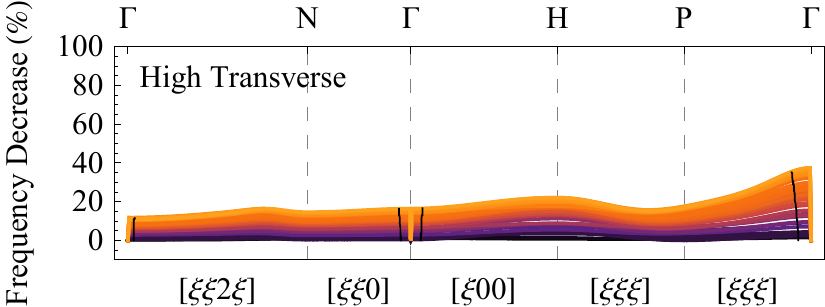}
\includegraphics[width=3.25in]{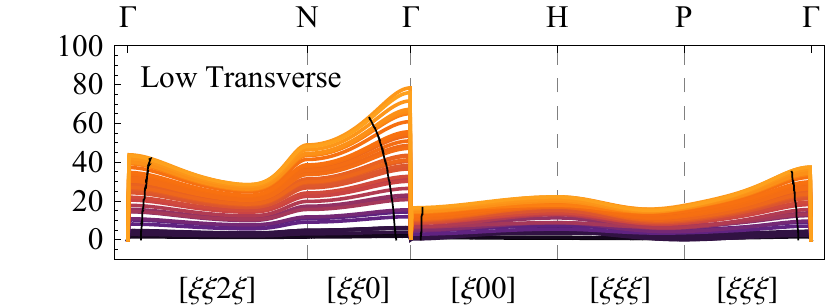}
\caption{\label{fig:DispersionPercentages} Percentage change in frequency with temperature of dispersions of Fig. \ref{fig:Dispersions}. The black lines denote the low energy dispersion regions below the dashed line in Fig. \ref{fig:Dispersions}. The colors correspond to temperatures as labeled in Fig. \ref{fig:DOS}.
}
\end{figure}
Figure \ref{fig:DispersionPercentages} shows the changes in the different dispersion branches with temperature relative to the 30K frequencies.
The measured phonon DOS spectra exhibited a 19\% decrease in phonon frequencies at the highest temperatures.
Figure \ref{fig:DispersionPercentages} makes clear that this average decrease is not evenly distributed across all the phonon modes. Most of the nonharmonic softening of the phonon DOS appears to originate in a few regions of the Brillouin zone.
Most notably, anomalously large softening is observed in the low transverse modes along the $\Gamma$-N direction, where all phonons soften by more than twice the average decrease observed in the phonon DOS.
The low transverse phonon branch $\rm T_{2}[\xi\xi0]$ corresponds to the $[1\overline{1}0]$ phonon polarization direction and softens significantly with temperature, consistent with the limited number of phonon dispersion studies on Fe at elevated temperature\cite{neuhaus_phonon_1997}. 
Large softening also occurs for the $[\xi\xi2\xi]$ branch and between the H and P high symmetry points at the 2/3 L [$\xi$,$\xi$,$\xi$] mode. 
Thermal softening seems to increase near the $\Gamma$ point on several high symmetry branches, but this increased softening at low-$q$ may be an artifact of the fitting method. 
Extracting a phonon DOS from the NRIXS spectra includes a removal of the elastic peak centered at zero energy transfer, requiring an extrapolation be used at energies below 4 meV (marked by the horizontal dotted line in Fig \ref{fig:Dispersions}). 
Accordingly, Fig. \ref{fig:DispersionPercentages} has black lines that delimit the low-q region corresponding to the elastic peak extrapolation, below which our fits are less reliable.

\begin{figure}[b]
\includegraphics[width=3.3in]{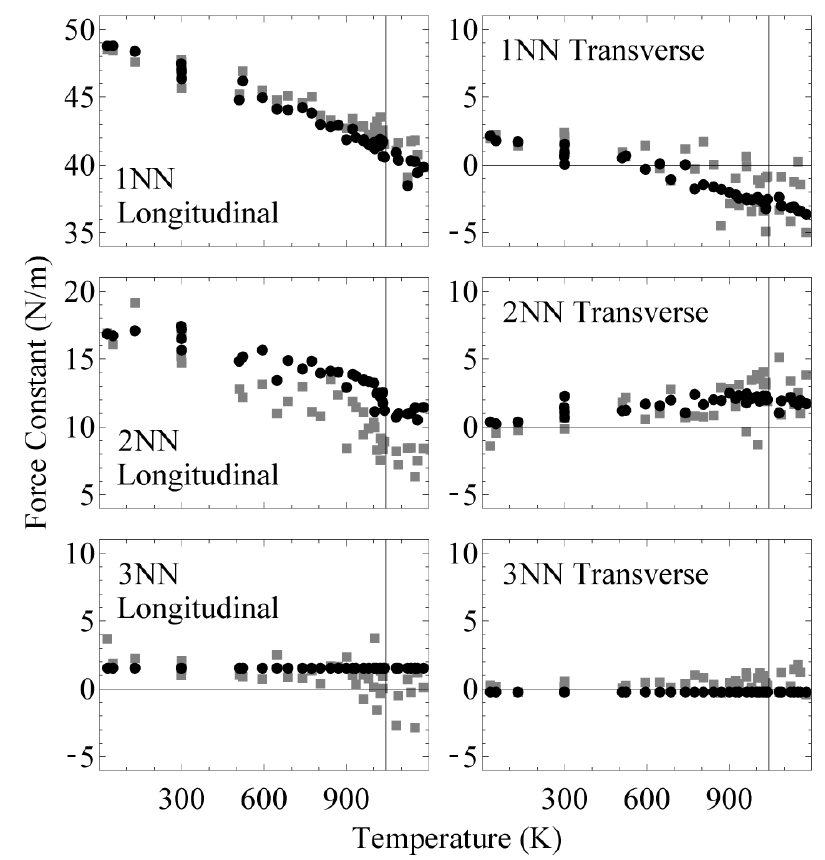}
\caption{\label{fig:ForceConstants} Extracted force constants from BvK fits. The force constant results using up through 4NN (11 variables) are shown in gray squares and those resulting from using up through 2NN (4 variables) are shown in black points. 
}
\end{figure}
\begin{figure}[b]
\includegraphics[width=3.3in]{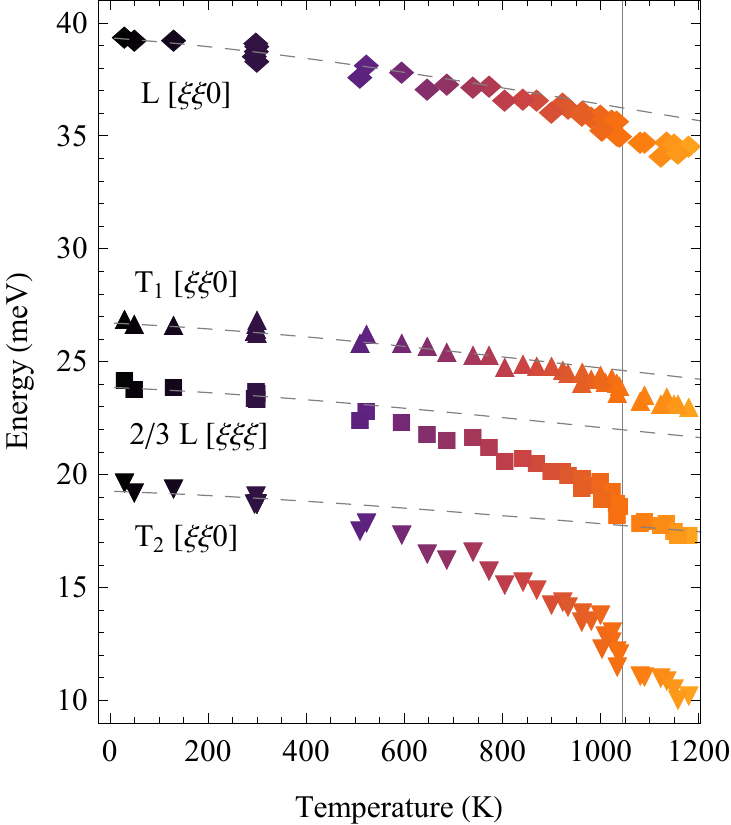}
\caption{\label{fig:DispersionPOI}Energies of specific phonon dispersion modes at the temperatures measured. The modes are compared with their quasiharmonic estimates (grey dashed lines). The colors correspond to temperatures as labeled in Fig. \ref{fig:DOS}.
}
\end{figure}
By projecting each nearest-neighbor tensorial component along the NN bond direction, axial and transverse force constants for bcc Fe were obtained for each nearest-neighbor pair as a function of temperature, as shown in Fig. \ref{fig:ForceConstants}.
With increasing temperature there is a large reduction in the first-nearest-neighbor (1NN) and second-nearest-neighbor (2NN) longitudinal force constants. 
The 2NN longitudinal force constant undergoes an especially strong softening. 
Above the Curie temperature the magnitude of the 2NN longitudinal force constant is reduced to 60\% (2NN fits) or 40\% (4NN fits) of its low temperature value. The 1NN longitudinal force constant decreases by only 20\% in the same temperature range.

The transverse force constants are calculated as an average of the tensorial force constants projected onto two vectors orthogonal to the bond direction. 
The values of these force constants are small and their trends are less reliable than for the larger longitudinal forces.
The magnitude of the 1NN average transverse force constant decreases rapidly at elevated temperatures and becomes negative beyond 800K, indicating a weakness of the bcc structure to shear stress\cite{petry_dynamical_1995,petry_phonons_1989,heiming_are_1991}. 
The 2NN transverse force constant appears to increase modestly with temperature, but this could be a compensation for changes in longer-range interatomic forces that were not varied in the fitting procedure. 
The inclusion of additional variables produced the same general trends as those displayed in Fig. \ref{fig:ForceConstants}, however, but with considerably more scatter.  
The large decrease in the 1NN and 2NN longitudinal force constants with temperature occurred for every optimization configuration that was tried.

\section{\label{sec:level1}Discussion}

\subsection{\label{sec:level2}Phonons and Born-von K\'{a}rm\'{a}n Model Dispersions}
There is significant phonon softening in bcc Fe at elevated temperatures.
A quasiharmonic model accounts for some of the measured phonon softening, but underestimates the thermal trends. 
Direct analysis of the phonon spectra shows that all the phonon DOS features exhibit softening beyond the prediction of the quasiharmonic model, and this excess softening is most notable in the low transverse modes. 
Both the departure from the quasiharmonic model at moderate temperatures and the differential mode softening are indicative of strongly nonharmonic behavior. 
However, the thermal broadening of the DOS features associated with phonon anharmonicity is small.

Phonon dispersions calculated from fitted force constants show that most of the nonharmonic softening occurs in low energy phonon branches, while most of the higher energy longitudinal phonons soften by an amount closer to that predicted by the quasiharmonic model. 
The temperature dependence of several phonon dispersion modes are shown in Fig. \ref{fig:DispersionPOI}.
The largest thermal softening is found for the low transverse $\rm T_{2}[\xi\xi0]$ branch. 
Anomalous softening of these phonons has been associated with dynamical precursors toward the fcc transition\cite{hasegawa_phonon_1987,petry_phonons_1989,heiming_are_1991,petry_dynamical_1995}. 
A combination of a displacement along the $\rm T_{2}[\xi\xi0]$ phonon mode, coupled with low-$q$ shearing consistent with $\rm T_{2}[\xi\xi0]$ and low-$q$ shearing along the $\rm T_{2}[\xi\xi2\xi]$ branch is a possible path for the structural transformation\cite{petry_phonons_1989}. 
All these modes soften anomalously with temperature. 
Softening of the modes on the Brillioun zone face between the H and P high symmetry points, most noticeably at 2/3 $\rm L[\xi\xi\xi]$, have been associated with the structural instability of the bcc lattice under pressure towards the hexagonal $\omega$-phase. 
The dynamical precursors to the $\alpha$-$\gamma$ transition in Fe seem to originate with the softening of the [$\xi$,$\xi$,0] branch, which is much larger than the softening of the [$\xi$,$\xi$,$\xi$] branch that is characteristic of the structural $\omega$-phase transition in the Group 4 bcc metals (Ti, Zr, Hf)\cite{petry_dynamical_1995,petry_phonons_1989} and Cr \cite{trampenau_temperature_1993} at elevated pressures. 
A large decrease in 2NN longitudinal forces was reported in bcc chromium at high temperatures, but Cr melts before the 2NN longitudinal force constant reaches the low values seen here for Fe\cite{trampenau_temperature_1993}.
The soft phonons shown in Fig. \ref{fig:DispersionPOI} begin to deviate from quasiharmonic behavior several hundred degrees below the magnetic transition, and continue to soften above the Curie temperature.
This anomalous phonon softening occurs in the same temperature range as the rapid decrease in the magnetization of $\alpha$-Fe\cite{crangle_magnetization_1971}.
Magnetic short range order has long been suspected of being important for the phonon thermodynamics of the paramagnetic phase\cite{holden_magnetic_1984},
and recent DFT calculations that account for paramagnetic interactions have successfully predicted the phonon dynamics at these temperatures\cite{kormann_atomic_2012}.

Simulations were performed to vary the longitudinal and traverse force constants individually for each nearest-neighbor pair.
Adjusting the 1NN longitudinal force constant with the others fixed had no effect on the $\rm T_{2}[\xi\xi0]$ branch.
Decreasing the 2NN longitudinal force constant relative to the others resulted in a rapidly softening $\rm T_{2}[\xi\xi0]$ branch that made the system dynamically unstable (imaginary phonon frequencies) when this force constant dropped below zero. 
When the 1NN transverse force constant decreased below -5 N/m, the phonons of the $\Gamma$-N branch also became dynamically unstable. 

A strong decrease of the 2NN longitudinal force constant with temperature can produce the significant non-harmonic softening observed in the $\Gamma$-$N$ phonon branch\cite{hasegawa_phonon_1987}. 
Similarly, anomalous behavior was seen in the 2NN magnetic exchange interaction parameters in a detailed study of magnon-phonon coupling \cite{yin_effect_2012}. 
It was found that including vibrational effects (including local volume and orientation) had a strong effect on the magnetic exchange interaction of the nearest-neighbor pairs, most notably for the second nearest-neighbors. 
The abnormal 2NN magnetic exchange behavior may be linked to the anomalous 2NN mode softening seen here. 

\subsection{\label{sec:level2}Vibrational Entropy and Free Energy}

\begin{figure}[b]
\includegraphics[width=3.3in]{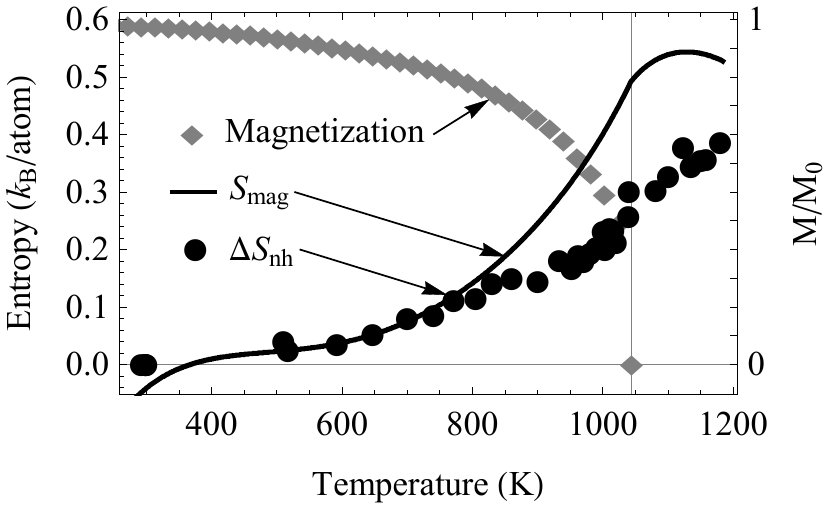}
\caption{\label{fig:AnharmonicVibEntropy} The nonharmonic vibrational entropy, $\Delta S_{\rm nh}$ from measured phonon DOS spectra, compared to the magnetization of bcc Fe\cite{crangle_magnetization_1971}, and the magnetic vibrational entropy, $S_{\rm mag}$, obtained by subtracting $S_{\rm vib}$ and $S_{\rm el}$\cite{jacobs_thermodynamic_2010} from the SGTE total entropy, $S$\cite{dinsdale_sgte_1991}.
}
\end{figure}

Our measurements of the phonon DOS spectra over a range of temperatures permits the direct assessment of vibrational entropy and vibrational free energy of bcc $\alpha$-Fe. 
The vibrational entropy from NRIXS measurements increases faster than predicted by the quasiharmonic model, $\Delta S_{\rm qh} + S_{\rm h}$. 
Any linear trends extracted from our force constants do not coincide with the volume normalized values provided by Klotz, et al., for bcc Fe under pressure\cite{klotz_phonon_2000}. 
The purely volume-dependent (quasiharmonic) effect from measurements at elevated pressure are quite different from our measured nonharmonic effects at high temperature. 
Furthermore, linear fits to our tensorial force constants are not capable of accurately reproducing the temperature dependence of the measured vibrational entropy of Fe.
There is a noticeable disagreement on either side of the magnetic transition; the force constants have a nonlinear thermal trend through the Curie temperature. 

The discrepancy between the quasiharmonic vibrational entropy, $\Delta S_{\rm qh} + S_{\rm h}$, and the measured vibrational entropy, $S_{\rm vib}$, is the nonharmonic entropy contribution, $\Delta S_{\rm nh}$.
Figure \ref{fig:AnharmonicVibEntropy} shows that the vibrational entropy of $\alpha$-Fe has a significant nonharmonic contribution, $\Delta S_{\rm nh}$.
The nonharmonic vibrational entropy, $\Delta S_{\rm nh}$, is compared to the magnetic entropy, $S_{\rm mag}$, and the magnetization also shown in Fig. \ref{fig:AnharmonicVibEntropy}. 
$S_{\rm mag}$ was calculated by subtracting our vibrational entropy, $S_{\rm vib}$, and also the electronic contribution, $S_{\rm el}$, as described by Jacobs, et al.,\cite{jacobs_thermodynamic_2010} from the total entropy of the SGTE database, $S$\cite{dinsdale_sgte_1991}. 
At temperatures just below the $\alpha$-$\gamma$ phase transition, $\Delta S_{\rm nh}$ changes the free energy by about 35meV/atom.


\begin{figure}[b]
\includegraphics[width=3.3in]{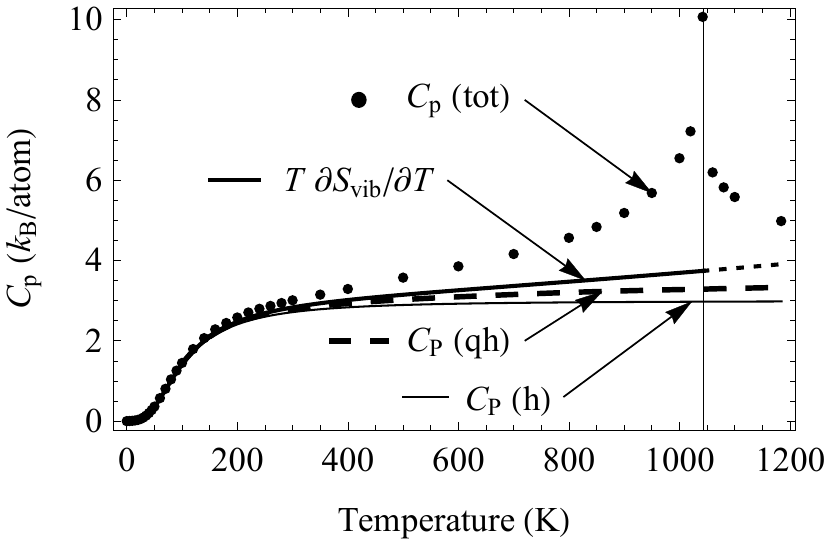}
\caption{\label{fig:HeatCapacity} The heat capacity calculated from fits to our vibrational entropy measurements (solid line), compared with the measured heat capacity from White, et al.\cite{white_thermophysical_1997} in points, heat capacity from the harmonic model (thin line), and heat capacity from the quasiharmonic estimate (dashed line).}
\end{figure}

Measurements of heat capacity at constant pressure, $C_{\rm P}$, have provided some of the most important experimental information on the thermodynamics of $\alpha$-Fe\cite{weiss_components_1956,kormann_free_2008,jacobs_thermodynamic_2010,yin_effect_2012,white_thermophysical_1997}, 
and the contributions from vibrational models are compared in Fig. \ref{fig:HeatCapacity}. 
Because heat capacity is obtained as a derivative quantity of the phonon entropy, we present heat capacity curves obtained from polynomial fits to our experimental $\Delta S_{\rm nh}$ results.
As such, the curves in Fig. \ref{fig:HeatCapacity} should be reliable for gradual trends, but possible features near the Curie temperature may be missing.
Nevertheless, it is clear that the $\Delta S_{\rm nh}$ of Fig. \ref{fig:AnharmonicVibEntropy} makes a significant contribution to the heat capacity, larger than the usual quasiharmonic contribution also shown in Fig. \ref{fig:HeatCapacity}.

The nonharmonic vibrational entropy can be written as 
\begin{equation}
\label{eq:NonHarmonicComponents}
\Delta S_{\rm nh}(T) = S_{\rm ppi}(T)+S_{\rm epi}(T)+S_{\rm mpi}(T),
\end{equation}
which includes the vibrational entropy from phonon-phonon interactions, $S_{\rm ppi}(T)$, vibrational entropy from electron-phonon interactions, $S_{\rm epi}(T)$, and vibrational entropy from magnon-phonon interactions, $S_{\rm mpi}(T)$.
Experimental measurements of phonon DOS spectra cannot alone be used to identify the individual terms $S_{\rm ppi}$, $S_{\rm epi}$ or $S_{\rm mpi}$ of $\Delta S_{\rm nh}$.
Nevertheless, the thermal trends are suggestive.
The $S_{\rm ppi}$ contribution from phonon-phonon interactions (often called the ``anharmonic'' contribution), arises from both the cubic and quartic components of the phonon potential \cite{wallace_thermodynamics_1998}.
With the cubic contribution comes a lifetime broadening from the imaginary part of the phonon self energy.
Even at the highest temperature, the lifetime broadening of the phonon DOS in $\alpha$-Fe is small compared to other systems \cite{tang_anharmonicity-induced_2010,tang_first-principles_2011,kresch_neutron_2007,kresch_phonons_2008,kresch_temperature_2009}.
For a damped harmonic oscillator, the shift in phonon frequencies, $\Delta$, associated with lifetime broadening is $\Delta = -\Gamma^2 / (2 \varepsilon)$, where $\Gamma$ is the linewidth broadening and $\varepsilon$ is the oscillator mode energy \cite{fultz_vibrational_2010}.
The broadening of the DOS features was assessed by examining the widths of the Lorentzian fits used to create Fig. \ref{fig:ModeTracking}.
From this measured broadening, the classical anharmonic shift, $\Delta$, is at least an order of magnitude smaller than the observed high temperature shifts.
Finally, phonon-phonon interactions from both cubic and quartic perturbations increase linearly with temperature, and the $\Delta S_{\rm nh}$ in Fig. \ref{fig:AnharmonicVibEntropy} does not follow a linear trend.
It appears that phonon-phonon interactions are not the main contribution to $\Delta S_{\rm nh}$ at high temperatures.

Electron-phonon coupling has been investigated by spin-polarized DFT calculations \cite{verstraete_ab_2013}, and effects were found to be modest.
These calculations did find large differences in the electron-phonon interactions for the majority and minority spin electrons, but did not consider disordered spin configurations.
Second-nearest-neighbor magnetic exchange interactions were reported to be anomalously sensitive to local atomic configurations \cite{yin_effect_2012}, and we found that 2NN force constants decrease significantly at temperatures where the spin order was decreasing rapidly.
The $\Delta S_{\rm nh}$ curve has a strikingly similar shape to the magnetic entropy curve in Fig. \ref{fig:AnharmonicVibEntropy}.
The magnon dispersions in iron have a maximum energy approximately an order of magnitude higher than the phonon dispersions \cite{boothroyd_high_1992}, but perhaps 5\% of the magnons are in the  energy range of phonons in Fe.
More processes involving two magnons may affect the phonon self energies.
A detailed analysis of phonon-magnon interactions is required for further progress, but it seems plausible that $S_{\rm mpi}$ is large.

\begin{figure}[b]
\includegraphics[width=3.3in]{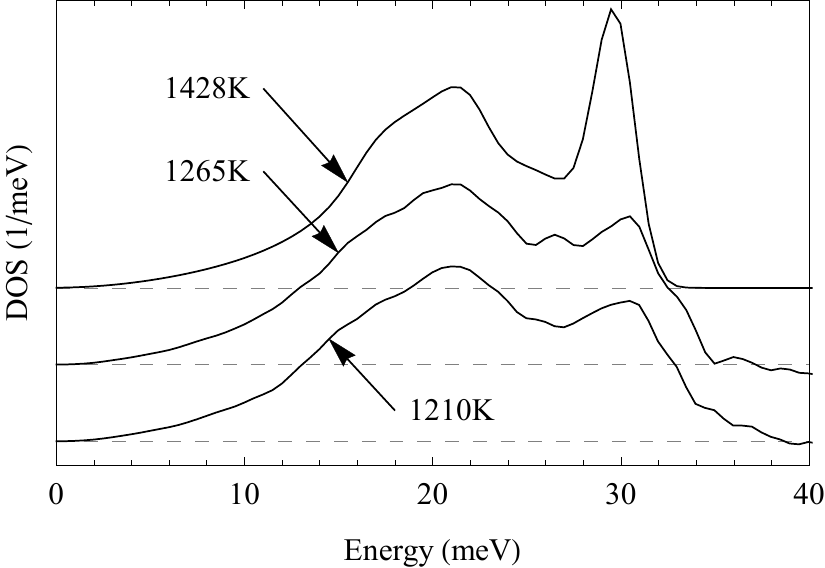}
\caption{\label{fig:FCC}The phonon DOS at two temperatures in the fcc phase, compared with a spectrum from neutron triple axis measurements\cite{zarestky_lattice_1987}.}
\end{figure}

Two phonon DOS spectra were acquired when the sample was in the fcc phase above 1185K, and these are shown in Fig. \ref{fig:FCC}. 
From these measurements, the change in vibrational entropy across the $\alpha$-$\gamma$ phase transition at 1185K was found to be 0.05$\,k_{\rm B}$/atom. 
This is notably smaller than previous literature values of 0.091 and 0.14$\,k_{\rm B}$/atom \cite{neuhaus_phonon_1997}, which are similar to the SGTE recommended value of 0.103$\,k_{\rm B}$/atom\cite{dinsdale_sgte_1991}, but the latter also includes magnetic and electronic contributions.
The fcc $\gamma$-Fe DOS spectra of Fig. \ref{fig:FCC} are compared with a 1428K $\gamma$-Fe DOS, calculated using force constants from the literature\cite{zarestky_lattice_1987} and convolved with our experimental NRIXS resolution for comparison.
The mean energies of the three DOS spectra are quite similar, but the two NRIXS DOS spectra measured here are significantly broader than the DOS calculated from dispersion measurements.
High temperature phonon measurements of several fcc metals exhibited significant phonon lifetime broading effects owing to phonon-phonon interactions, so these effects may make a more significant contribution to the thermodynamics of fcc $\gamma$-Fe \cite{tang_anharmonicity-induced_2010,tang_first-principles_2011,kresch_neutron_2007,kresch_phonons_2008,kresch_temperature_2009}.
A more systematic study of the shape of the phonon DOS in the fcc phase should help determine if there is a large lifetime broadening, but if so we may have underestimated the vibrational entropy for the fcc phase\cite{palumbo_thermodynamic_2014}.

\section{Conclusions}

Nuclear resonant inelastic x-ray scattering was used to measure the phonon DOS of bcc $\alpha$-Fe from low temperature up through the  $\alpha$-$\gamma$ transition.
The vibrational entropy deviated significantly from predictions of quasiharmonic theory by as much as 0.35$k_{\rm B}$/atom (or a free energy contribution \SI{35}{\milli\electronvolt}/atom) at 1150K.
The nonharmonic contribution $\Delta S_{\rm nh}$ was distinctly nonlinear with temperature, and occurred without significant broadening of the phonon lineshape, unlike typical behavior with phonon-phonon interactions.
The temperature-dependence of $\Delta S_{\rm nh}$ followed the magnetic entropy, however, suggesting that the change of magnon-phonon interactions with temperature makes a significant contribution to the nonharmonic phonon softening of $\alpha$-Fe.
The vibrational entropy of the bcc-fcc Fe transition at 1185\,K was found to be smaller than the assessed thermodynamic value.

A Born--von K\'{a}rm\'{a}n model was fit to the experimental phonon DOS spectra, and used to extract interatomic force constants.
Full phonon dispersions were then calculated from the Born--von K\'{a}rm\'{a}n force constants.
These dispersions showed that the anomalous softening originates primarly from low transverse modes along the $\Gamma$-N high symmetry direction, in agreement with single crystal triple axis neutron studies.
The anomalous softening originates with the large softening of the 2NN longitudinal force constant, which may be consistent with the atypical sensitivity of the 2NN exchange interaction to local atomic configurations.

\begin{acknowledgments}

This work benefitted from important discussions with Drs. W. Sturhahn, F. K\"{o}rmann, B. Grabowski, T. Hickel, and J. Neugebauer. 
This work was supported by the Department of Energy through the Carnegie-DOE Alliance Center, funded by the Department of Energy
through the Stewardship Sciences Academic Alliance Program.
This work benefited from DANSE software developed under NSF Grant No. DMR-0520547.
Portions of this work were performed at HPCAT (Sector 16), Advanced Photon Source (APS), Argonne National Laboratory.  HPCAT is supported by CIW, CDAC, UNLV and LLNL through funding from DOE-NNSA, DOE-BES and NSF. Use of the APS was supported by DOE-BES, under Contract No. DE-AC02-06CH11357.

\end{acknowledgments}


\bibliography{HighTFe_Manuscript}

\subsection{\label{sec:level1}Appendix}

\begin{figure}[b]
\includegraphics[width=3.in]{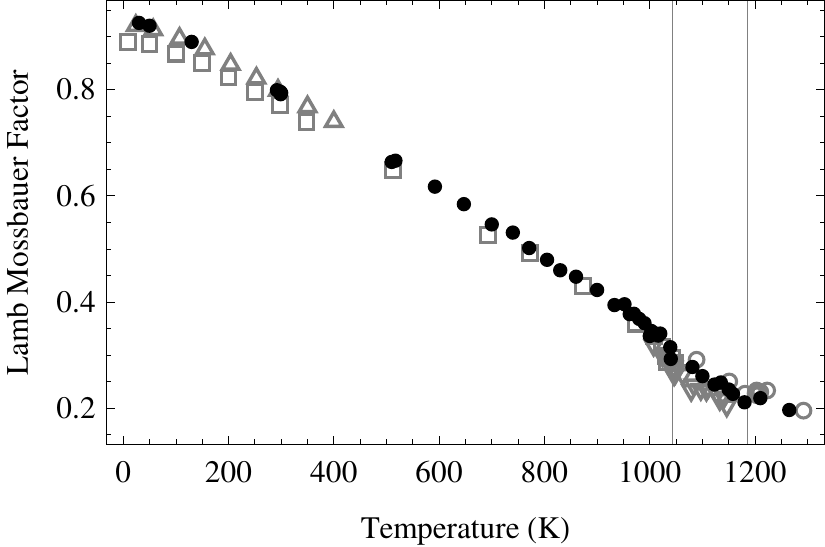}
\caption{\label{fig:LambMossabuer} 
Lamb-M\"{o}ssbauer factors calculated from measured NRIXS spectra. The experimental data from this study (presented in black) are compared with literature values in open circles \cite{kovats_mossbauer_1969} open squares \cite{bergmann_temperature_1994} open up triangles \cite{chumakov_temperature_1996} and open down triangles \cite{kolk_recoilless_1986}. 
}
\end{figure}

\end{document}